\documentclass[journal]{IEEEtran}

\usepackage{verbatim}
\usepackage{amsfonts}
\usepackage{amssymb}
\usepackage{stfloats}
\usepackage{cite}
\usepackage{graphicx}
\usepackage{psfrag}
\usepackage{amsmath}
\usepackage{array}
\usepackage{epstopdf}
\usepackage{authblk}
\usepackage{graphicx} 
\usepackage{amsthm} 
\usepackage{lipsum}
\usepackage{verbatim} 
\usepackage{authblk}
\usepackage{mathtools}
\usepackage{cuted}
\usepackage{booktabs}

\usepackage{amsmath}
\usepackage{mathrsfs}

\usepackage{algorithmic}

\usepackage{array}

\ifCLASSOPTIONcompsoc
\usepackage[caption=false,font=normalsize,labelfont=sf,textfont=sf]{subfig}
\else
\usepackage[caption=false,font=footnotesize]{subfig}
\fi

\usepackage{url}
\usepackage{graphicx,amsmath,amssymb,amsfonts}
\usepackage{algorithmic,algorithm}

\usepackage{hyperref}
\hypersetup{colorlinks=true, citecolor=blue}

\usepackage{setspace}

\makeatletter

\newcommand{\Rmnum}[1]{\expandafter\@slowromancap\romannumeral #1@}
\makeatother

\usepackage{framed} 

\usepackage{cancel}

\usepackage{flushend}

\usepackage{booktabs}
\usepackage{tabularx,makecell,multirow}

\usepackage[table]{xcolor}
\usepackage{pifont} 

\usepackage{diagbox}

\begin{document}
	% reference control
	\bstctlcite{ref:BSTcontrol}

	\title{From Single to Multi-Functional RIS: Architecture, Key Technologies, Challenges, and Applications}
	\author{Wanli~Ni, %\IEEEmembership{Member,~IEEE},
		Ailing~Zheng, Wen~Wang,
		Dusit~Niyato, %\IEEEmembership{Fellow,~IEEE},
		Naofal Al-Dhahir, %\IEEEmembership{Fellow,~IEEE},
		and Mérouane Debbah %\IEEEmembership{Fellow,~IEEE}
	\vspace{-1 mm}
	\thanks{Wanli Ni is with the Department of Electronic Engineering, Tsinghua University, China (e-mail: niwanli@tsinghua.edu.cn).}
	\thanks{Ailing Zheng and Wen Wang are with the State Key Laboratory of Networking and Switching Technology, Beijing University of Posts and Telecommunications, China (e-mail: ailing.zheng@bupt.edu.cn, wen.wang@bupt.edu.cn).}
	\thanks{Dusit Niyato is with the College of Computing and Data Science, Nanyang Technological University, Singapore (email: dniyato@ntu.edu.sg).}
	\thanks{Naofal Al-Dhahir is with the Department of Electrical and Computer Engineering, The University of Texas at Dallas, USA (email: aldhahir@utdallas.edu).}
	\thanks{Mérouane Debbah is with the KU 6G Research Center, Khalifa University of Science and Technology, UAE (e-mail: merouane.debbah@ku.ac.ae).}
}
	
	\maketitle
	
	\begin{abstract}
		Although reconfigurable intelligent surfaces (RISs) have demonstrated the potential to boost network capacity and expand coverage by adjusting their electromagnetic properties, existing RIS architectures have certain limitations, such as double-fading attenuation and restricted half-space coverage.
		In this article, we delve into the progressive development from single to multi-functional RIS (MF-RIS) that enables simultaneous signal amplification, reflection, and refraction.
		We begin by detailing the hardware design and signal model that distinguish MF-RIS from traditional RISs.
		Subsequently, we introduce the key technologies underpinning MF-RIS-aided communications, along with the fundamental issues and challenges inherent to its deployment.
		We then outline the promising applications of MF-RIS in the realm of communication, sensing, and computation systems, highlighting its transformative impact on these domains.
		Lastly, we present simulation results to demonstrate the superiority of MF-RIS in enhancing network performance in terms of spectral efficiency.
	\end{abstract}

%	\vspace{-3 mm}
	\section{Introduction}
	Cutting-edge applications such as holographic communications, virtual reality, and self-driving cars are likely to be the killer applications in the upcoming 6G network \cite{Wang20236G}.
	However, current resource-constrained wireless communication systems are struggling to meet the increasing demand for quality of user experience, including high data rates, low latency, and ubiquitous connectivity, posed by these advanced applications.
	Moreover, with the widespread implementation of millimeter-wave (mmWave) bandwidth, future networks may face a number of challenges, such as severe path loss, beam misalignment, and coverage holes.
	Against this background, reconfigurable intelligent surfaces (RIS) stands out as a potential technology to address the aforementioned issues.
	Specifically, the RIS can effectively enhance the coverage, spectral and energy efficiency of existing wireless networks by regulating the radio propagation environment in a cost-effective and environment-friendly manner \cite{Ni2021Resource}.
	
\begin{figure*}[t]
		\centering
		\includegraphics[width=7 in]{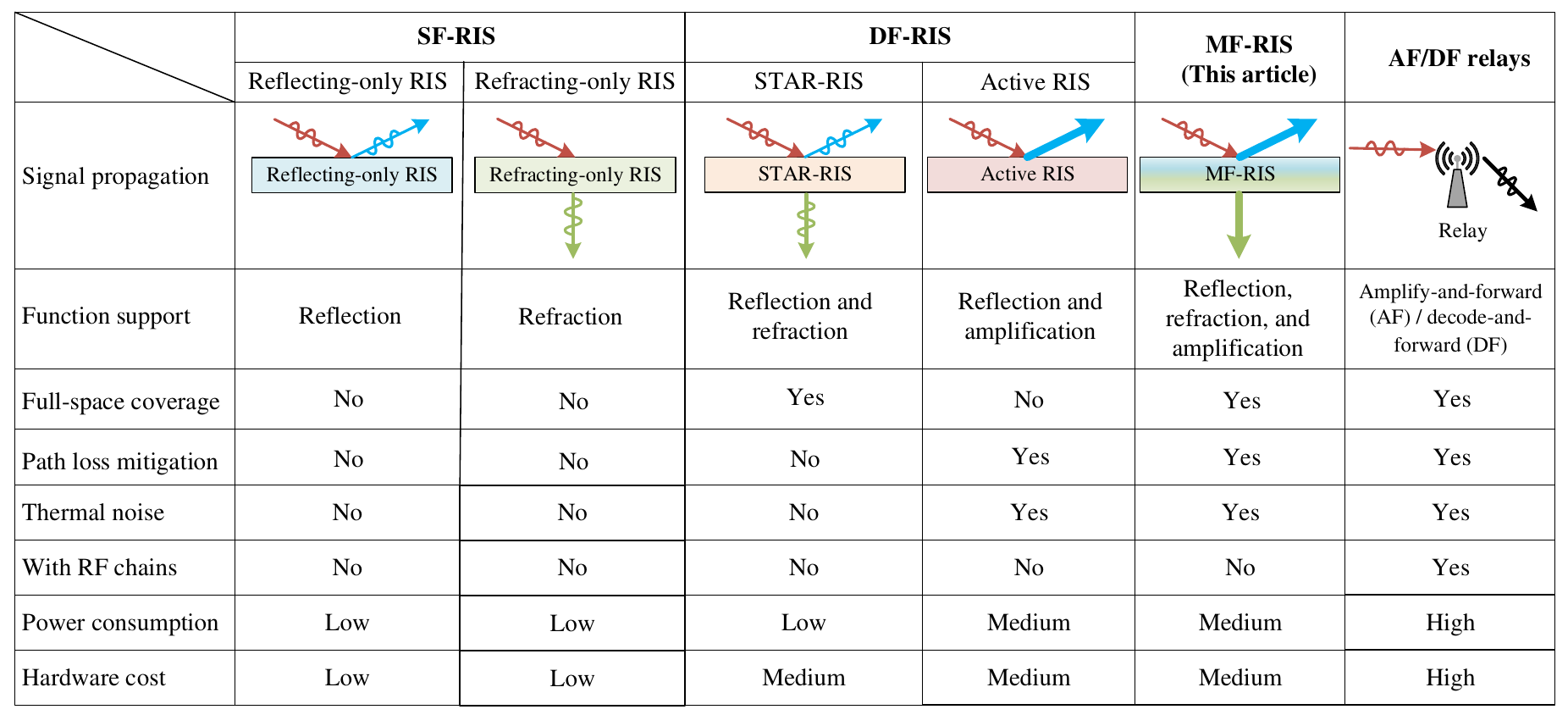}
		\caption{MF-RIS versus traditional SF-RIS, DF-RIS, and relays.}
		\label{Fig1}
	\end{figure*}

	Currently, the majority of research efforts have been concentrated on single-functional RIS (SF-RIS), which supports either passive reflection or refraction \cite{Ni2021Resource, Wu2020Beamforming, Shao2024Intelligent}.
	However, SF-RIS suffers from a topological constraint, limiting its service to users located in a specific half-space.
	This limitation significantly hinders its deployment flexibility and efficiency in wireless networks with randomly distributed users.
	To address this issue, recent studies have explored dual-functional RIS (DF-RIS), such as simultaneous transmission and reflection RIS (STAR-RIS) \cite{Liu2021STAR}, which promises full-space coverage and an omni-dimensional smart radio environment.
	Although DF-RIS overcomes the half-space coverage limitation, both SF-RIS and DF-RIS face a common issue, i.e., signals relayed through them undergo double-fading attenuation due to the cascaded base station (BS)-RIS-user link \cite{Wu2020Beamforming}.
	In this article, we propose the new concept of multi-functional RIS (MF-RIS), which is capable of realizing three functions on one meta-surface.
	Remarkably, MF-RIS not only achieves full-space coverage by integrating signal reflection and refraction, but also mitigates the double-fading issue through signal amplification \cite{Wang2023IoT, Zheng2023WCL, Wang2024MFRIS}.
	This innovative RIS has the potential to revolutionize the performance and flexibility of future RIS-aided networks.	
	Furthermore, in Fig.~\ref{Fig1}, we compare MF-RIS to traditional RISs and relays to distinguish them clearly.
	
	Above, we have provided justification for why we need MF-RISs.
	Now, we aim to delve deeper into three fundamental questions:
	1) What is an MF-RIS?
	2) What are the key issues and challenges faced by the MF-RIS?
	3) What are the potential applications of the MF-RIS?
	To answer these questions, our responses are summarized as follows.
	\begin{itemize}
		\item
		\textbf{What is an MF-RIS:}
		Our MF-RIS comprises an array of active elements having a three-layer structure, which integrates signal reflection, refraction, and amplification functions simultaneously.
		This allows MF-RIS to overcome limitations of traditional RISs, such as half-space coverage and double-fading attenuation.
		The hardware architecture and key technologies are given in Sections~\ref{Section2} and \ref{Section3}, respectively.

		\item
		\textbf{What are the key issues and challenges of MF-RIS:}
		When deploying MF-RIS in wireless networks, several key issues and challenges include the optimization under non-ideal conditions (e.g., discrete and coupled coefficients), the selection of static and dynamic RISs, two-timescale beamforming, and interference management of distributed RISs.
		See Section \ref{Section4} for more details.
		
		\item
		\textbf{What are the potential applications of MF-RIS:}
		The versatility of MF-RIS can be incorporated into diverse scenarios to enhance coverage, security, connectivity, and energy efficiency, facilitating novel services and applications such as wireless sensing, computing, and energy transfer.
		In Section \ref{Section5}, we outline seven promising directions to guide the implementation of MF-RIS.
% Since MF-RIS can be deployed in diverse scenarios to improve coverage, security, connectivity, spectral and energy efficiency, MF-RIS can enable new services and applications in 6G networks like wireless sensing, computing, and power transfer.
	\end{itemize}

%	\newpage
	\section{Next-Generation RIS: From Single to Multiple Functions} \label{Section2}
	By integrating signal reflection, refraction, and amplification functions into one meta-surface, we present the next-generation RIS, termed MF-RIS.
	Unlike existing RISs, the MF-RIS is able to control multiple electromagnetic wave parameters in a wider range, so as to realize more complex communication functions, which is beneficial to meet different user requirements.
	In the following, we present the hardware architecture of MF-RIS and its corresponding signal model.

	\begin{figure*}[t]
		\centering
		\includegraphics[width=6.3 in]{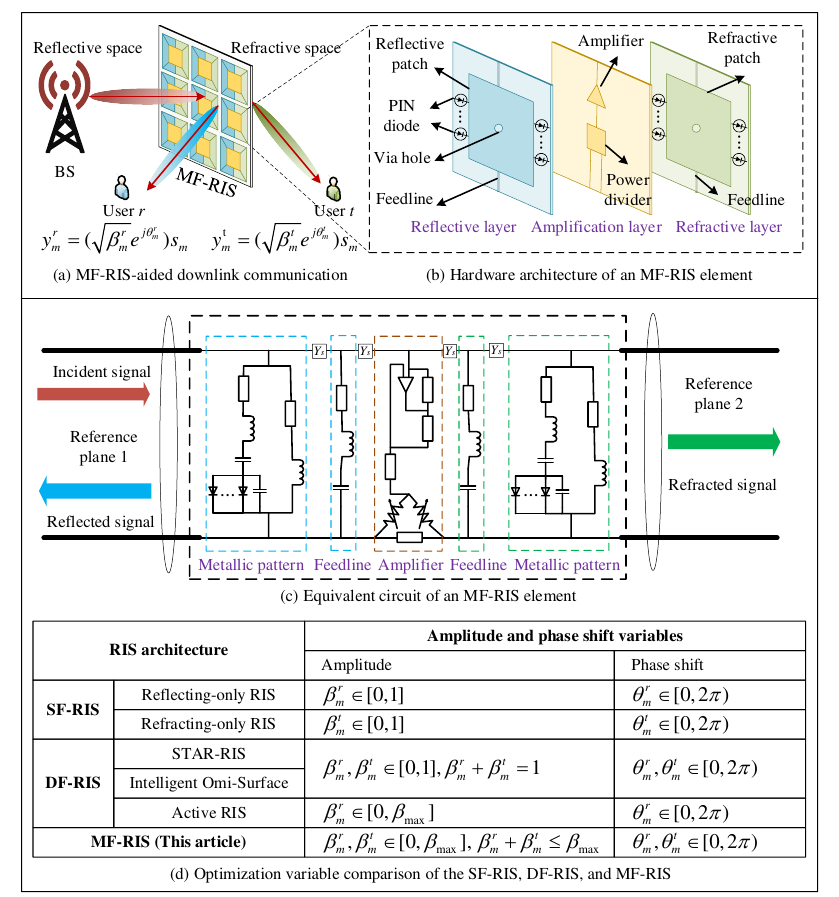}
		\caption{Hardware design and signal model of MF-RIS: (a) an MF-RIS-aided downlink communication system, (b) hardware architecture of an MF-RIS element, (c) equivalent circuit of an MF-RIS element, and (d) optimization variable comparison of the SF-RIS, DF-RIS, and MF-RIS.}
		\label{Fig2}
	\end{figure*}

	\subsection{Hardware Architecture}
%	By integrating signal reflection, refraction, and amplification functions into one meta-surface, the MF-RIS is able to overcome the drawbacks such as half-space coverage and double-fading attenuation faced by other RIS counterparts. 
	As shown in Fig. \ref{Fig2}, we design a novel hardware architecture for the proposed MF-RIS which comprises an array of active elements.
	Each element features a three-layer structure (i.e., reflective, amplification, and refractive) that enables multi-dimensional and full-space signal manipulation \cite{Wang2023IoT}.
	Specifically, the reflective and refractive layers are designed symmetrically to facilitate simultaneous signal reflection and refraction. Each of them consists of a metallic patch, a feedline, and multiple reconfigurable components.
	The amplification layer houses an amplifier and a power divider that are used to dynamically adjust the amplitude of both reflected and refracted signals by supplying additional power \cite{Zheng2023WCL}.
	This design combines the benefits of passive RIS elements with active components, enhancing signal control flexibility.
	
	Compared with conventional RISs, amplify-and-forward (AF) and decode-and-forward (DF) relays, the MF-RIS offers the following benefits for practical implementation.
	\begin{itemize}
		\item 
		\textbf{Phase and amplitude control with amplification:}
		Different from conventional RISs, the MF-RIS offers precise control over both the amplitude and phase of transmitted signals, allowing for sophisticated beamforming and spatial signal processing. This is crucial for high-performance communications and advanced applications in future wireless networks.
		\item 
		\textbf{Full-duplex operation without reception:}
		Unlike traditional relays, the MF-RIS amplifies incident signals directly without the need for separate reception and transmission stages. This distinction not only simplifies signal processing but also enables full-duplex operation without requiring complex self-interference cancellation techniques, improving spectral efficiency while reducing the implementation cost and complexity.
	\end{itemize}

	\subsection{Signal Model}
	To formulate the signal model of MF-RIS-aided communications, we consider an MF-RIS comprising $M$ elements.
	The reflective and refractive phase shifts of the $m$-th element are denoted by $\theta_m^r$ and $\theta_m^t \in [0, 2\pi)$, respectively.
	Similarly, the reflective and refractive amplitude coefficients are represented by $\beta_m^r$ and $\beta_m^t \in [0,\beta_{\max}]$, respectively, where $\beta_{\max} \ge 1$ is the maximum amplification power at each element \cite{Wang2024MFRIS}.
	As such, the reflective and refractive matrices of MF-RIS are modeled as
	$\Theta_r = {\rm diag}(\sqrt{\beta_1^r}e^{j\theta_1^r}, \sqrt{\beta_2^r}e^{j\theta_2^r},\dots,\sqrt{\beta_M^r}e^{j\theta_M^r})$ and
	$\Theta_t = {\rm diag}(\sqrt{\beta_1^t}e^{j\theta_1^t}, \sqrt{\beta_2^t}e^{j\theta_2^t},\dots,\sqrt{\beta_M^t}e^{j\theta_M^t})$, respectively.
	The signal received by the $m$-th element is denoted by $s_m$.
	Then, following the manipulation principle of the MF-RIS, the reflected and refracted signals transmitted by the MF-RIS can be modeled as
	$y_m^r = (\sqrt{\beta_m^r}e^{j\theta_m^r})s_m$ and
	$y_m^t = (\sqrt{\beta_m^t}e^{j\theta_m^t})s_m$, respectively.
	According to the law of energy conservation, 
	the amplitude coefficients should satisfy $\beta_m^r + \beta_m^t \le \beta_{\max}$, ensuring that the energy consumed by the amplifier does not exceed the maximum available energy at the MF-RIS \cite{Zheng2023CL}.
	
	In Fig. \ref{Fig2}, we also compare the parameter configurations of the conventional RIS and the novel MF-RIS.
	From a mathematical perspective, the MF-RIS encompasses SF-RIS and DF-RIS as special cases within its broader scope \cite{Wang2024MFRIS}. 
	Specifically, by setting $\beta_{\max} = 1$, the MF-RIS aligns itself with the characteristics of a STAR-RIS. Further narrowing these conditions to $\beta_m^t = 0$ and $\theta_m^t = 0$, the MF-RIS effectively mirrors a reflecting-only RIS.
	These reductions demonstrate the generality of the MF-RIS, highlighting its flexible and versatile design philosophy.

	\begin{figure*}[t]
		\centering
		\includegraphics[width=7 in]{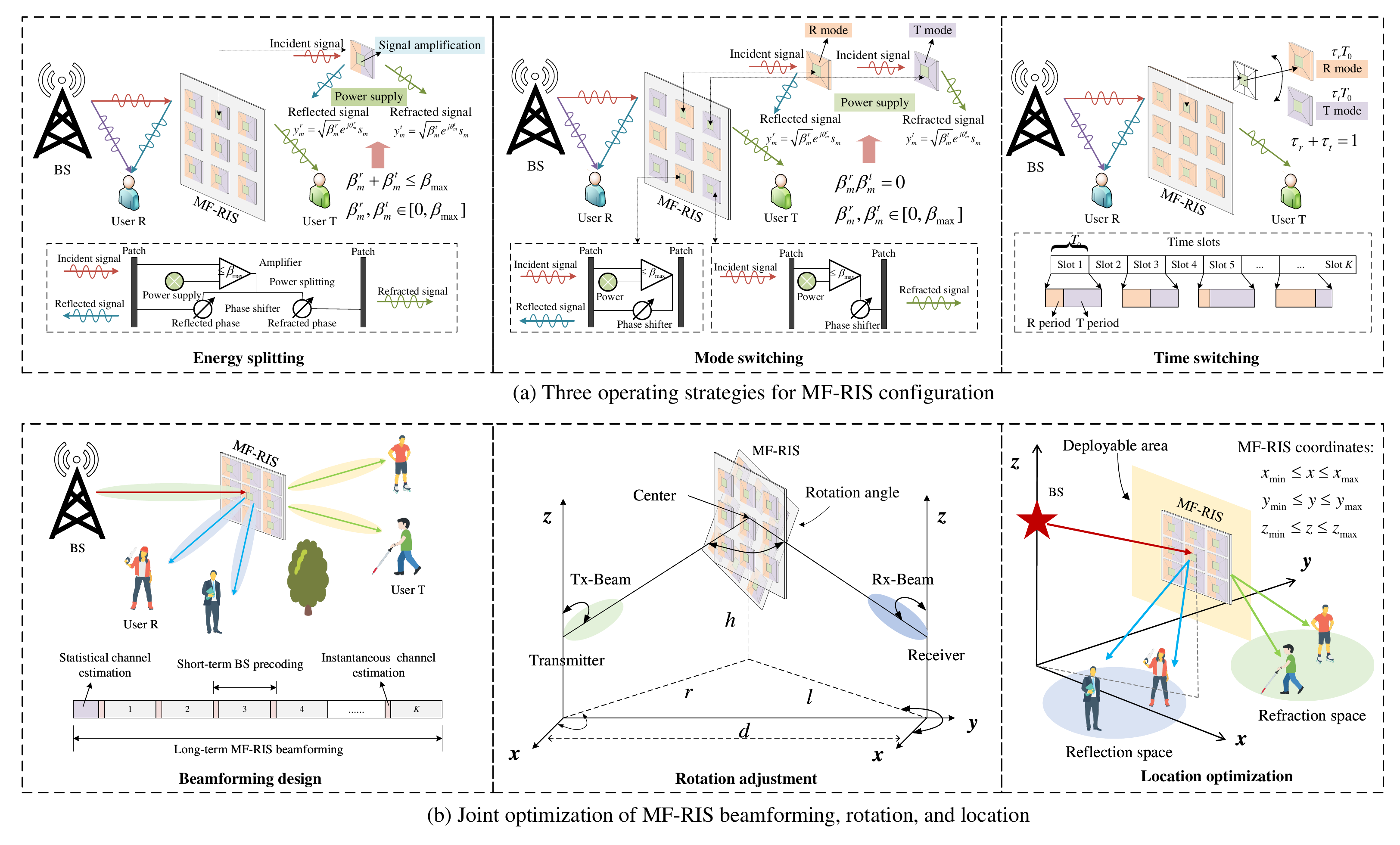}
		\caption{Key enabling technologies for MF-RIS-aided wireless communications: (a) three operating strategies for MF-RIS configuration, (b) joint optimization of MF-RIS beamforming, location, and rotation.}
		\label{Fig3}
	\end{figure*}
	
	\section{Key Enabling Technologies for MF-RIS} \label{Section3}
	In this section, we present some key enabling technologies for the practical deployment of MF-RIS in communication systems, including the operating strategy, location and rotation angle optimization, channel estimation and robust beamforming design of MF-RIS.
	
	\subsection{Operating Strategy}
	As illustrated in Fig. \ref{Fig3}, we present three operating strategies for MF-RIS configuration in wireless networks~\cite{Zheng2023CL}.
	
	\textbf{Energy splitting (ES):}
	All MF-RIS elements perform signal amplification, reflection, and refraction functions simultaneously.
	After the incident signal is amplified, its energy is split into reflection and refraction outputs.
	This strategy optimizes all variables but incurs high hardware complexity.

	\textbf{Mode switching (MS):}
	All MF-RIS elements are divided into two groups: the reflective and refractive groups.
	The former (latter) group is responsible for reflecting (refracting) the amplified signal.
	This strategy simplifies the hardware design but sacrifices some of the performance gain.
	
	\textbf{Time switching (TS):}
	All MF-RIS elements perform either signal reflection or refraction in one time slot.
	This strategy decouples the reflection and refraction coefficients and simplifies the optimization problem, but increases the time synchronization requirement.
	
	\subsection{Channel Estimation}
	While the MF-RIS can amplify signals effectively, it lacks the ability to transmit or receive pilot symbols for channel estimation.
	Therefore, acquiring CSI in MF-RIS-aided communication systems primarily relies on the transmitter and the receiver.
	Broadly, existing channel estimation methods for RISs can be categorized into two types: 1) estimating individual channels separately and 2) estimating the direct and cascaded channels jointly.
	In particular, traditional techniques such as least squares (LS) and minimum mean square error (MMSE) are often used to acquire CSI in a low-complexity manner.
	Due to the sparse nature of mmWave channels, compressive sensing has emerged as a reliable solution for channel estimation in RIS-aided mmWave networks.
	In addition, deep learning, known for its nonlinear mapping capabilities, offers a promising direction for learning approximate mapping functions from training data to predict CSI.
	During channel estimation, it is typically assumed that the MF-RIS does not perform amplification, allowing existing estimation methods developed for STAR-RIS to be adopted for MF-RIS-aided systems.
	
	\subsection{Robust Beamforming Design}
	Although MF-RIS shows great potential in improving system performance, it encounters significant difficulty in acquiring perfect CSI.
	This is mainly due to channel estimation errors and feedback delays brought by the large number of MF-RIS elements.
	Additionally, the built-in amplifiers in MF-RIS lead to increased thermal noise, further compromising the accuracy of the acquired CSI.
	Therefore, it is imperative  to design robust beamforming schemes in MF-RIS-aided wireless communication systems.
	To characterize the CSI uncertainty, bounded or statistical error models are commonly adopted.
	Specifically, for the bounded error model, mathematical tools including the S-procedure and general sign-definiteness approximations are utilized to tackle the semi-infinite constraints that arise due to CSI uncertainty.
	Meanwhile, for the statistical error model, statistical methods such as Bernstein-type inequalities are applied to optimize system parameters through safe approximations, thus ensuring stable performance even under imperfect CSI.

	\subsection{Joint Location and Rotation Optimization}
	Both the electromagnetic properties and directionality of MF-RIS significantly affect the overall system performance.
	Although previous studies on MF-RIS have made some progress in signal amplitude and phase adjustment, most of them are confined to fixed MF-RIS locations, failing to fully unleash the potential of MF-RIS.
	The 3D location and rotation angle of MF-RIS affect the channel quality between the transmitter and receiver.
	Therefore, employing advanced machine learning and optimization techniques to jointly optimize the location and rotation of MF-RIS can make full use of the additional spatial degrees of freedom (DoFs) offered by MF-RIS to mitigate the effects of signal fading, interference, and mobile users, which further promotes the performance boundaries of existing systems by increasing the opportunities of line-of-sight (LoS) communication.
	In summary, the precise adjustment of the 3D location and rotation angle of MF-RIS introduces novel optimization DoFs for the study of MF-RIS-aided communications.

	\begin{figure}[t]
		\centering
		\includegraphics[width=3.5 in]{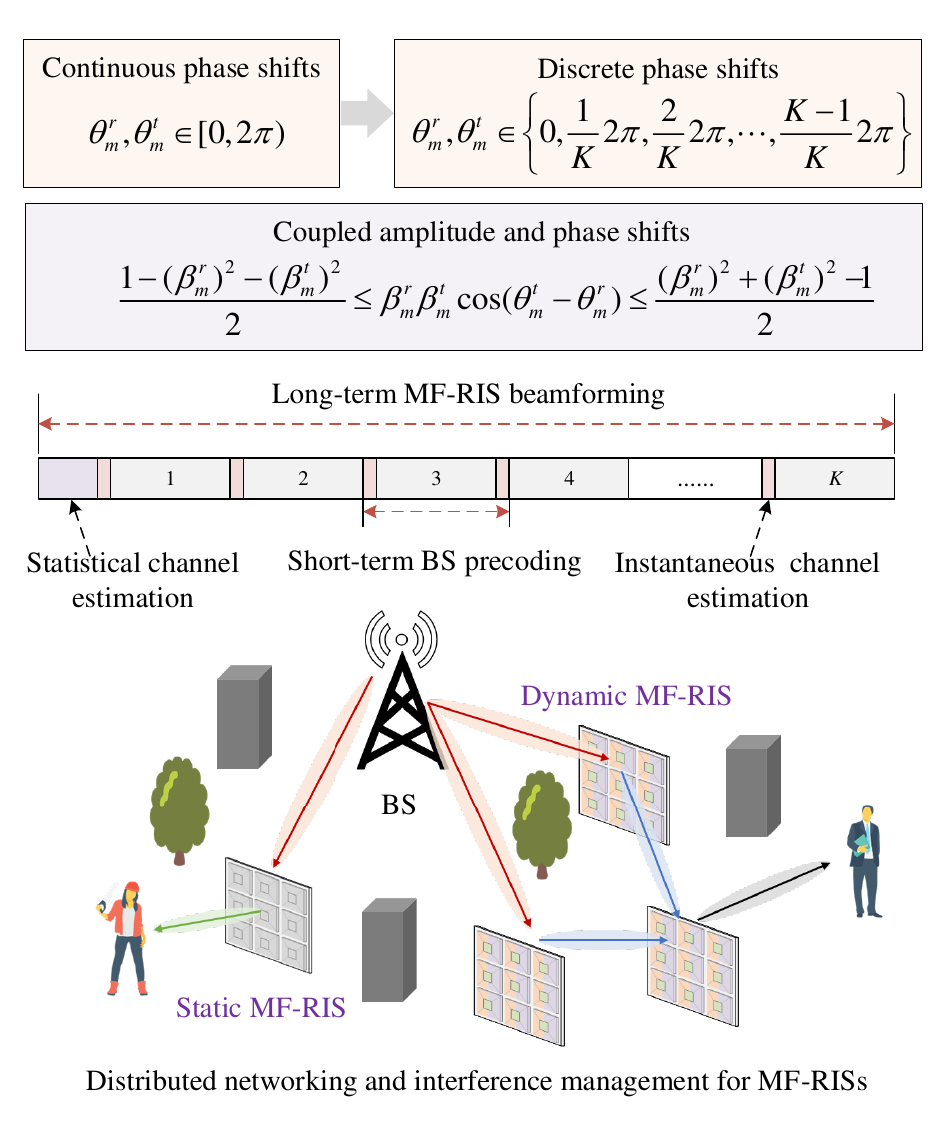}
		\caption{Important fundamental issues and challenges in MF-RIS-aided wireless communication systems.}
		\label{Fig4}
	\end{figure}
	
	\section{Fundamental Issues and Challenges of MF-RIS} \label{Section4}
	Due to imperfect physical implementation and inherent hardware limitations, the MF-RIS faces some fundamental issues and challenges, such as discrete phase shifts, coupled coefficients, two-timescale beamforming, and interference management, as shown in Fig. \ref{Fig4}.
	
	% Hardware Impairments
	\subsection{Discrete Phase Shifts}
	Most of the existing studies assume that the RIS phase shifts are continuous.
	However, in practice, due to physical implementation constraints, such as manufacturing cost and inherent hardware limitations, the phase shifts of MF-RIS are typically discrete rather than continuous \cite{Zheng2023WCL}.
	This hardware constraint introduces quantization errors, which subsequently affects the accuracy of MF-RIS beamforming.
	As a result, the signals transmitted through different links fail to be constructively combined at the receiver, ultimately degrading the overall system performance.
	In addition, increasing the phase shift resolution requires embedding more PINs at each element, leading to higher costs and scalability issues as the number of MF-RIS elements grows.
	This also creates additional challenges in the design of an RIS controller, as more complex circuitry is required to manage the increased number of PIN diodes.
	Therefore, it is crucial to explore an optimal phase shift configuration of low-resolution MF-RIS in upcoming 6G networks.
	
	The most straightforward approach to optimize discrete phase shifts is exhaustive search, but this is computationally infeasible when the number of MF-RIS elements is large \cite{Ni2022Integrating}.
	A more practical alternative is the approximation projection method.
	Here, the discrete phase shifts are initially relaxed to continuous values, enabling more applications of optimization techniques. Once the optimal continuous phase shifts are obtained, they are then quantized to their nearest discrete equivalents \cite{Zheng2023WCL}.
	Furthermore, heuristic alternating optimization techniques can be utilized to optimize the discrete phase shifts by iteratively fixing the values of all elements except one \cite{Wu2020Beamforming}. This approach provides a more efficient solution for large-scale MF-RIS-aided communication systems.

	%% 潜在的关键问题和技术挑战
%	\subsection{Nonlinear Amplifiers}
	
	\subsection{Coupled Amplitude and Phase Shifts}	
	When optimizing traditional SF-RIS, the amplitude and phase shift coefficients are often tuned independently.
	However, the MF-RIS operates differently, where its amplitude and phase shift coefficients are tightly-coupled \cite{Zheng2024TVT}.
	This is because each MF-RIS element concurrently supports electric and magnetic surface currents to enable signal reflection and refraction.
	These currents are governed by the element's electric and magnetic impedances, which are inherently interconnected rather than independent.
	This interconnection results in a coupling relationship between the phase shift and amplitude coefficients in MF-RIS, which also increases the complexity and difficulty of solving problems.
	Therefore, traditional methods of independently optimizing amplitude and phase shift is no longer applicable. It is crucial to devise innovative methodologies that can effectively address this coupling, as it is essential for enhancing the overall performance of MF-RIS-aided communication systems.
	
%	Traditional SF-RIS exhibits purely imaginary electric and magnetic impedances.
%	This characteristic allows for the independent tuning of the phase shift and amplitude coefficients.
%	However, for MF-RIS with active components, its impedances can assume values not greater than zero.
%	This results in a close coupling between the amplitude and phase shift coefficients of MF-RIS, necessitating a different optimization approach.
	
	Through theoretical analysis in \cite{Zheng2024TVT}, we reveal the coupling relationship between the electrical and magnetic impedance of MF-RIS and derive a closed-form expression for the coupled amplitude and phase shift model of MF-RIS.
	One potential approach to optimizing coupled amplitude and phase shift coefficients is to decouple these variables and optimize them separately by leveraging Euler's formula.
	From our previous work in \cite{Zheng2024TVT}, we obtain the following insights:
	1) As the amplification power increases, the coupling degree between the amplitude and phase shift of each element loosens, allowing for gradual independent adjustment of these parameters.
	2) When the MF-RIS operates in either pure refraction or reflection mode, the phase shift can be randomly configured without compromising optimality.

	\subsection{Static and Dynamic MF-RIS}
	By precisely adjusting multi-dimensional coefficients of MF-RIS by a controller, we can ensure that the radiation signals propagate in a specified direction precisely throughout the entire spatial domain.
	However, this advanced functionality also comes with an increase in hardware complexity, computational overhead, and energy consumption.
	To strike a balance between hardware costs, operational expenditures, and system performance across varying application scenarios and channel conditions, we offer two MF-RIS schemes:
	\begin{itemize}
		\item 
		\textbf{Static MF-RIS:} This type of MF-RIS employs low-complexity, non-adjustable hardware. Once deployed, its reflection, refraction, and amplification coefficients remain constant throughout its operational lifespan. This cost-effective MF-RIS scheme is particularly suitable for fixed, small-area coverage enhancement in quasi-static environments dominated by LoS channels.
		\item
		\textbf{Dynamic MF-RIS:} Unlike the static one, this MF-RIS type requires more complex, adjustable hardware materials. It offers the flexibility to dynamically adjust the reflection, refraction, and amplification coefficients for precise beam steering and user tracking in real-time. This MF-RIS scheme significantly enhances the coverage of the BS and ensures seamless connectivity for mobile users in a large area, making it suitable for dynamic environments with rich scattering channels.
	\end{itemize}

	\subsection{Two-Timescale Beamforming Design}
	Despite the immense potential of MF-RIS beamforming schemes leveraging instantaneous channel state information (CSI) to drastically boost wireless network performance, practical implementations encounter several key challenges:
	1) The BS tends to estimate the cascaded BS-RIS-User channel, overlooking the individual estimation of the RIS-User and RIS-BS channels \cite{Zhi2023Two}. 
	2) The vast number of elements in MF-RIS, coupled with channels on both sides, significantly inflates the pilot overhead for estimating individual links.
	3) Schemes relying on instantaneous CSI require frequent beamforming calculations and information feedback within each channel coherence interval, resulting in high computational complexity and large feedback overhead.
		
	To address these challenges, an innovative approach is to design a two-timescale beamforming scheme \cite{Zhi2023Two}. The core idea is to separate the optimization of BS precoding and MF-RIS beamforming into two distinct timescales:
	\begin{itemize}
		\item 
		\textbf{Short-term BS precoding:} Leveraging instantaneous CSI, a BS adjusts its precoding at millisecond intervals to swiftly adapt to dynamic channel variations.
		\item 
		\textbf{Long-term MF-RIS beamforming:} The MF-RIS updates its beamforming based on statistical CSI, typically on the order of seconds, to account for slower-changing channels.
	\end{itemize}
%	This two-timescale strategy not only ensures efficient joint optimization between BS precoding and MF-RIS beamforming but also grants greater flexibility in adapting to applications with varying time-sensitivity requirements. As a result, it significantly reduces channel estimation overhead, while mitigating computational complexity.

	\subsection{Distributed Networking and Interference Management}
	In complex environments such as stadiums and shopping malls, wireless signal propagation may encounter blockages.
	By deploying multiple MF-RISs at critical locations to enhance path diversity, we can leverage multi-hop signal reflection or refraction to provide high-quality communication services for ubiquitous users.
	This multi-hop cooperation mechanism facilitates the creation of stable unobstructed channels, enabling desired signals to bypass obstacles and find the optimal propagation path, thus providing beamforming gains that exceed traditional single-hop RIS-assisted communication.
	To maximize the performance of distributed MF-RIS deployment, we need to overcome the following key challenges:
	\begin{itemize}
		\item 
		\textbf{MF-RIS networking optimization:} 
%		According to users' needs and network conditions, 
		Selecting an optimal set of RISs for each user to maximize communication capacity is a critical issue that directly affects the networking efficiency and signal quality.
		\item 
		\textbf{Beam interference management:} In multi-user scenarios with distributed MF-RISs, signals may be relayed by multiple adjacent MF-RISs, which may cause beam interference and affect the overall performance.
%		For example, MF-RIS selected by different users should avoid LoS links between each other.
		\item 
		\textbf{Computing and hardware requirements:} Controlling multiple distributed MF-RISs simultaneously places higher demands on the computing capability of the BS and the hardware complexity of the RIS controller.
	\end{itemize}

%	\subsection{Increased Power Consumption}

	\begin{figure*}[t]
		\centering
		\includegraphics[width=7 in]{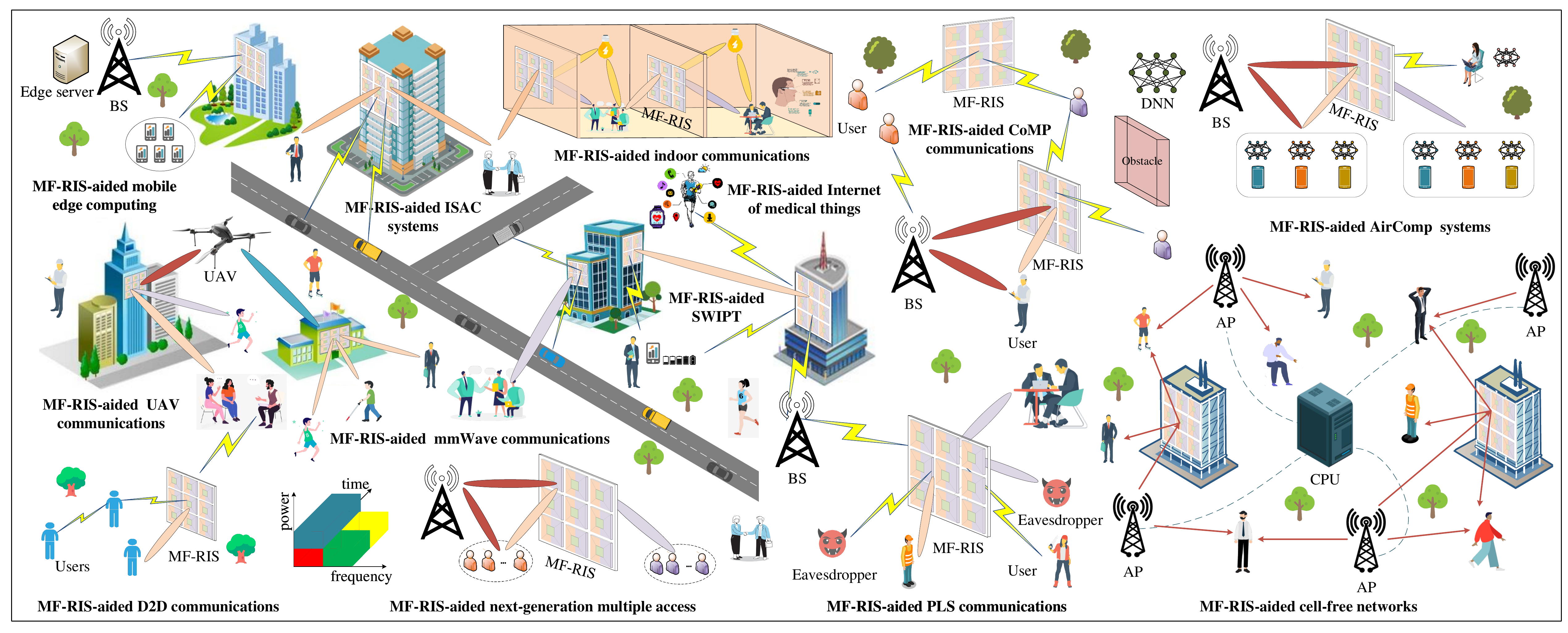}
		\caption{Illustration of some potential applications of MF-RIS in wireless communication, sensing, and computation systems.}
		\label{Fig5}
	\end{figure*}
	
	\section{Applications of MF-RIS in Communication, Sensing, and Computation Systems} \label{Section5}
	By empowering each element with multiple functions, the MF-RIS offers additional DoFs for signal manipulation, including enhancing the desired signal constructively and mitigating harmful interference destructively.
	In the following, we discuss several emerging applications of MF-RIS in communication, sensing, and computation systems.
	More potential applications of MF-RIS are illustrated in Fig.~\ref{Fig5}
	
	\subsection{MF-RIS-Aided mmWave Communications}
	By virtue of its vast available bandwidth, mmWave communication has been considered a highly promising technology for enabling high-speed data transmission.
	However, the shorter wavelength of mmWave communications poses significant challenges in terms of severe path loss.
	Although deploying large antenna arrays can enhance the array gain and alleviate this issue to some extent, it introduces complexities in hardware design and imposes considerable power consumption.
	Moreover, the directional transmission of mmWave signals renders them highly susceptible to obstruction by objects such as humans, vehicles, and buildings.
	In this context, the deployment of MF-RIS as auxiliary transmission links for mmWave communication becomes paramount, especially when direct links are obstructed.
	The MF-RIS-aided mmWave communications not only avoids the need for large antenna arrays at the transmitter, but also further reduces overall power consumption and simplifies hardware design at the transceiver.
	
	\subsection{MF-RIS-Aided PLS Communications}
	Physical layer security (PLS) leverages the inherent physical characteristics of wireless channels to achieve secure communication through advanced techniques such as artificial noise, cooperative interference, and secure beamforming.
	However, under adverse wireless conditions, particularly in scenarios with multiple eavesdroppers and multi-path interference, PLS methods may become ineffective.
	To address this challenge, we employ MF-RIS to create a secure radio environment.
	By adjusting beamforming and power allocation strategies, the MF-RIS is able to boost the signal quality received by legitimate users while effectively suppressing the illegal signals eavesdropped by malicious users, thereby greatly enhancing communication security.
	Furthermore, MF-RIS can generate artificial noise to further degrade the information reception capabilities of eavesdroppers \cite{Wang2023UAV}.

%	\subsection{MF-RIS-Aided Movable Antenna Systems}	

	\subsection{MF-RIS-Aided Non-Terrestrial Networks}
	With their flexible deployment and efficient LoS communication, the non-terrestrial networks integrating unmanned aerial vehicles (UAVs) and high altitude platforms (HAPs) aim to achieve global coverage and ubiquitous connectivity, surpassing the limitations of traditional terrestrial networks. 
	Especially in rural areas and disaster scenarios, non-terrestrial networks are seen as a promising solution to expand the coverage of existing ground communication networks.
	However, factors such as the Doppler effect, power limitations, and long-distance transmission still hinder the performance of non-terrestrial networks.
	To overcome these challenges, we propose utilizing MF-RIS to establish relay links and effectively compensate for the Doppler shift in non-terrestrial networks, thus enhancing the quality of air-to-ground communication links.
	
	\subsection{MF-RIS-Aided Cell-Free Networks}
	Compared to traditional cellular networks, cell-free networks completely abandon the concept of cell boundaries by using a large number of geo-distributed access points (APs) to serve users collaboratively.
	This architecture significantly mitigates the negative impact of shadow fading caused by obstacles such as buildings and terrain on communication quality.
	Moreover, users can readily connect with nearby APs, achieving shorter communication distances and thus reducing transmission losses.
	However, the extensive deployment of numerous APs also poses challenges, such as increased cost and energy consumption.
	To deal with this issue, an MF-RIS-aided cell-free network was proposed in \cite{Wei2024Multi}.
	The simulation results in \cite{Wei2024Multi} demonstrated that, compared with the benchmark without RIS, a throughput enhancement of 74.0\% can be obtained if a 60-element MF-RIS is employed, while achieving higher energy efficiency.

	\subsection{MF-RIS-Aided SWIPT Systems}
	Simultaneous wireless information and power transfer (SWIPT) utilizes radio frequency (RF) signals to transmit both information and energy concurrently, extending the operational lifespan of energy-constrained devices.
	However, SWIPT faces several challenges:
	1) Large-scale channel fading affects communication stability and wireless energy transfer efficiency.
	2) The path loss of RF signals limits the effective deployment range of energy users.
	To address these issues, MF-RIS can be integrated into SWIPT systems to establish stable reliable links, enhancing communication stability and reducing the negative impact of channel fading.
	Additionally, MF-RIS amplifies incoming signals, effectively expanding the signal coverage area and enabling more devices to receive sufficient energy and information.
	This improves energy utilization efficiency and reduces overall energy consumption.
	
%	\subsection{MF-RIS-Aided Wireless Sensing}

	\subsection{MF-RIS-Aided ISAC Systems}
	The integrated sensing and communication (ISAC) system aims to transmit dual-functional waveforms through a unified hardware platform to meet the needs of emerging applications such as autonomous robots and intelligent transportation for high-quality wireless communication and precise sensing.
	However, in the practical deployment of ISAC systems, several challenges arise, including the unavailability of LoS links, the absence of multi-path propagation, and the limited sensing range due to high path loss \cite{Shao2024Intelligent}.
	Integrating MF-RIS in ISAC systems can bring the following advantages.
	Firstly, MF-RIS can assist in constructing virtual LoS links, enabling the detection of targets even in the presence of obstacles.
	Secondly, by introducing additional DoFs and transmitting multi-source signals, MF-RIS creates a rich multi-path propagation environment, resulting in more accurate user positioning.
	Lastly, MF-RIS enables flexible beam scanning functionality to compensate for path loss.
	In summary, MF-RIS in ISAC systems not only reconfigures the radio propagation environment to enhance communication capacity but also significantly improves the performance of existing wireless sensing.

	\begin{figure*} [t!]
	%	\centering
		\subfloat[Sum-rate versus the total power budget]{\label{Fig6_1}
			\begin{minipage}[t]{0.45 \textwidth}
				\centering
				\includegraphics[width= 3.35 in]{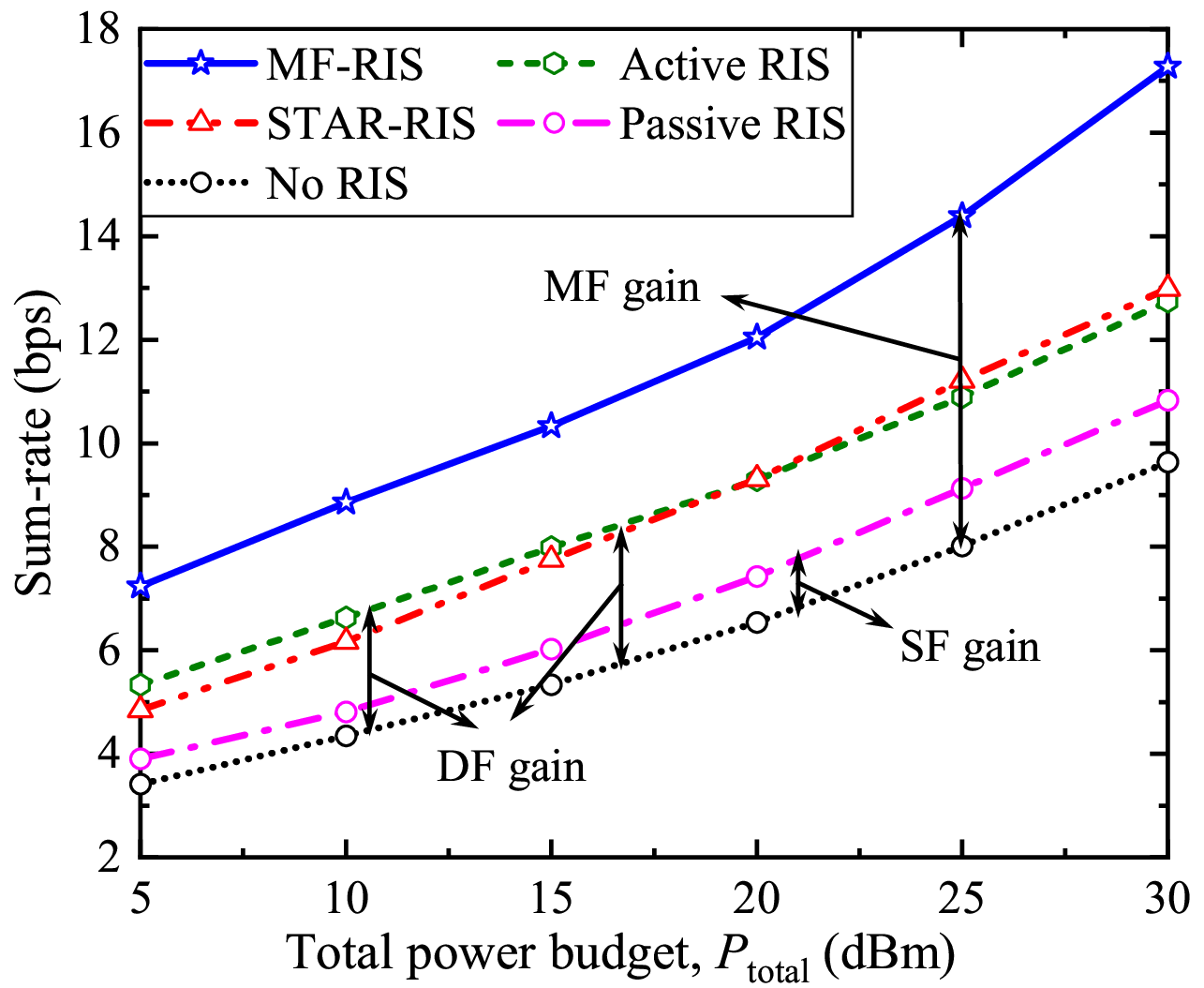}
			\end{minipage}
		}  \hspace{5 mm}
		\subfloat[Sum-rate versus the number of RIS element]{\label{Fig6_2}
			\begin{minipage}[t]{0.45 \textwidth}
				\centering
				\includegraphics[width= 3.35 in]{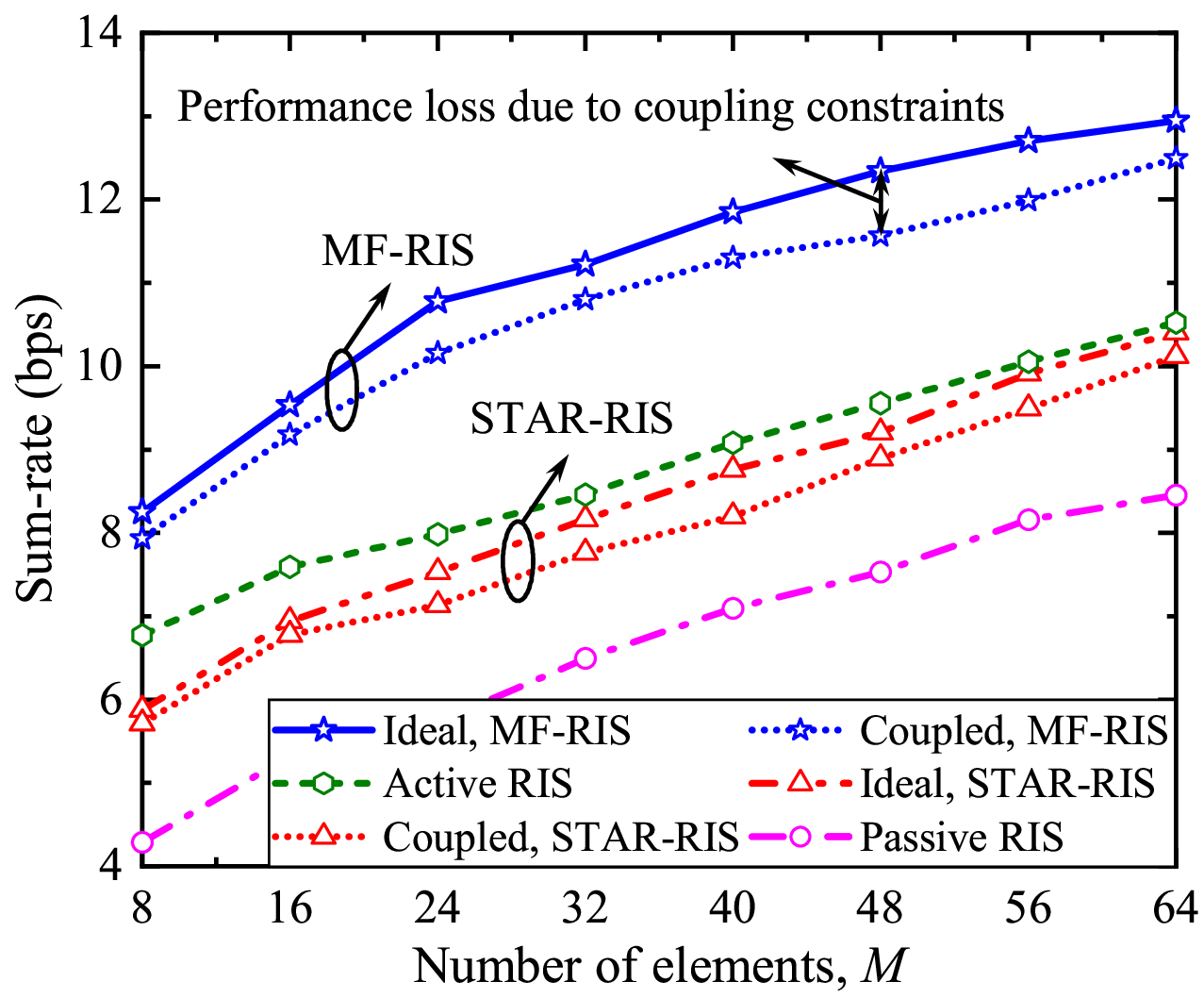}
			\end{minipage}
		}
		\caption{Performance evaluation: (a) sum-rate versus the total power budget, and (b) sum-rate versus the number of RIS element.}
		\label{learning_performance}
	\end{figure*}

	\subsection{MF-RIS-Aided AirComp Systems}
	Traditional data aggregation in IoT networks follows a {communicate-then-compute} paradigm, which requires the AP to collect data from each IoT device sequentially and then perform computation (such as averaging).
	In contrast, over-the-air computation (AirComp) fuses communication and computing, creating a new paradigm of {compute-while-communicate}.
	It allows all IoT devices to transmit data simultaneously over a shared wireless channel and exploits the waveform superposition property of multi-access channels for data aggregation~\cite{Ni2022STAR}.
	However, the performance of AirComp is greatly affected by the quality of wireless channels.
	Although passive RIS can mitigate adverse effects of poor channel conditions to some extent, it requires users to be located within a designated half space, and the RIS-aided cascaded channel experiences high path loss.
	Unlike passive RIS, MF-RIS not only achieves full-space data aggregation through signal reflection and refraction, but also mitigates the multiplicative fading effect by amplifying signals.
	As a result, the MF-RIS-aided AirComp system, with its precise calibration and enhancement of signals, can significantly reduce the mean squared error caused by signal superposition, and thereby achieving high-performance wireless data aggregation.
	
%	\subsection{MF-RIS-Aided Mobile Edge Computing}

	%	\begin{figure*}[t]
	%		\centering
	%		\includegraphics[width=3.5 in]{Fig6.pdf}
	%		\caption{Performance comparison: (a) sum rate vs. the transmit power, (b) sum rate vs. the number of elements, and (c) energy efficiency vs. the power budget.}
	%		\label{RIS_P}
	%	\end{figure*}

%	\newpage
	\section{Use Case}
	In this section, we present a case study to show the performance of MF-RIS-aided communications.
	Specifically, we assume that a BS equipped with $N=8$ antennas communicates with two users with the assistance of an MF-RIS comprising $M=64$ elements.
	For ease of presentation, we assume that two users are located on either side of the MF-RIS.
	The noise power is set to $-80 {\rm dBm}$.
	Let $P_{\mathrm{BS}}$ and $P_{\mathrm{RIS}}$ denote the transmit power at the BS and the amplification power at the MF-RIS, respectively.
	To make the comparison fair, the total available power for our scheme and the benchmarks is set to $P_{\mathrm{total}} = 20~\mathrm{dBm}$.
	Other parameters are similar to those in \cite{Zheng2023WCL} and \cite{Zheng2024TVT}, where an alternating optimization algorithm is utilized to jointly optimize the BS beamforming and MF-RIS coefficients.
	We compare MF-RIS with three existing RIS types: passive RIS, STAR-RIS, and active RIS.

	Fig. \ref{Fig6_1} depicts the sum-rate versus the total power budget.
	We observe that the proposed MF-RIS exhibits the best performance, followed by the two DF-RISs (active RIS and STAR-RIS), while the passive RIS, a type of SF-RIS, performs the worst. 
	This is because DF-RIS can only partially address the challenges faced by SF-RIS, while MF-RIS achieves full-space signal enhancement by integrating multiple functions on a single metasurface.
	Another interesting observation is that as $P_{\rm total}$ increases, the active RIS performance gradually surpasses that of STAR-RIS. 
	This indicates that when power is sufficient, signal amplification effectively enhances the quality of cascade links, outperforming simultaneous reflection and refraction in boosting system throughout.
	
	Fig. \ref{Fig6_2} shows the sum-rate versus the number of RIS elements.
	As expected, all curves exhibit increasing trends as $M$ grows up.
	However, as $M$ further increases, the performance growth rate of MF-RIS and active RIS gradually slows down compared to STAR-RIS. 
	This deceleration is attributed to the fact that, given the limited total power, the  power constraints of MF-RIS and active RIS become increasingly restrictive as $M$ rises.
	This suggests that their power allocation needs to be carefully balanced to meet the power consumption requirements of both the BS and the RIS.
	However, even when $M=64$, STAR-RIS is still inferior to MF-RIS when $M=32$.
	In addition, due to the coefficient coupling constraints, the more practical coupled MF-RIS is slightly inferior to the ideal MF-RIS.
	Nevertheless, it still outperforms other types of RIS, once again confirming the superiority of the proposed MF-RIS.

%	\newpage	
	\section{Conclusions}
	In this article, we proposed a groundbreaking MF-RIS to create a full-space smart radio environment and mitigate the double-fading attenuation encountered by existing RISs.
	Firstly, we introduced the hardware architecture and basic signal model of MF-RIS.
	Subsequently, we described the key enabling technologies for MF-RIS-aided communication systems, including operation and deployment strategies.
	Furthermore, we discussed major fundamental issues and challenges of MF-RIS, such as discrete phase shifts, coupled coefficients, and distributed networking.
	In addition, we outlined potential applications of MF-RIS in emerging wireless communication, sensing, and computation systems.
	Finally, we presented a use case to illustrate the efficacy and superiority of our MF-RIS compared to existing RIS architectures.

	%\newpage
	\bibliographystyle{IEEEtran}
	\bibliography{IEEEabrv,ref}

\end{document}